\documentclass[prl,twocolumn,showpacs]{revtex4}% Physical Review B

\usepackage[dvips]{graphicx}% Include figure files
\usepackage{dcolumn}% Align table columns on decimal point
\usepackage{bm}% bold math
\usepackage{amssymb}

\DeclareGraphicsExtensions{eps}

\begin{document}
\preprint{APS/123-QED}
\title{Spin Anisotropy and Slow Dynamics in Spin Glasses}% Force line breaks with \\
\author{F. Bert}
\altaffiliation[Present address: ]{%
Laboratoire de Physique des Solides, Universit\'{e} Paris-Sud,
B\^at. 510, 91405 Orsay, France}%
\author{V. Dupuis}
\altaffiliation[Present address: ]{%
Laboratoire des Milieux D\'esordonn\'es et H\'et\'erog\`enes, Universit\'{e} Paris 6,
Paris, France}%
\author{E. Vincent}
\author{J. Hammann}%
\author{J.-P. Bouchaud}
\affiliation{Service de Physique de l'\'Etat
Condens\'{e}, DSM, CEA Saclay,
91191 Gif-sur-Yvette Cedex, France.}%Lines break automatically or can be forced with \\

\date{\today}% It is always \today, today,

\begin{abstract}
We report on an extensive study of the influence of spin 
anisotropy on spin glass aging dynamics.  
New temperature cycle experiments allow us to compare 
quantitatively the memory effect in four Heisenberg spin 
glasses with various degrees of random anisotropy and one Ising spin glass.  
The sharpness of the memory effect appears to 
decrease continuously with the spin anisotropy. Besides, the spin glass 
coherence length is determined by magnetic field change experiments for the 
first time in the Ising sample. For three representative samples, from Heisenberg to 
Ising spin glasses, we can consistently account for both sets of experiments (temperature cycle 
and magnetic field change) using a single expression for the growth of the coherence length with
time.
\end{abstract}

\maketitle

Real spin glasses (SGs), which are both disordered and frustrated magnetic systems, 
are always found to be out of equilibrium on the laboratory time scale. 
When cooled below its freezing temperature $T_g$, 
a SG starts to age; from a completely random initial configuration, spin-spin 
correlations develop and progressively `rigidify' the spin network.  
If the SG is further cooled, the aging process restarts (rejuvenation effect) 
but the previous aging is not lost. When reheated, the SG retrieves its former state 
(memory effect) and resumes its former evolution. These memory and rejuvenation effects 
suggest a description in terms of a hierarchical organization of metastable states as 
a function of the temperature~\cite{Vincent97}, although an
alternative description in terms of temperature chaos has also been proposed~\cite{Yoshino01}. 
Recently, a comparison between an Heisenberg and an Ising SG showed that the slow dynamic 
properties are quantitatively different in these systems~\cite{Dupuis01}, the memory effect 
being more pronounced in the Heisenberg case. How are the spin anisotropy and the aging 
properties related? Is there a systematic effect of the anisotropy? Answering these questions 
would shed a new light on the longstanding and topical problem of the nature of the 
Heisenberg SG phase~\cite{Lee03}. The relevance of anisotropy for this problem is supported 
by the recent finding that the critical exponents vary continuously with the SG 
anisotropy~\cite{Petit02}.

We report in this letter on two classes of experiments. 
First, we have performed 
thermo-remanent magnetization (TRM) measurements with negative temperature cycles 
during aging in one Ising and several Heisenberg SGs in order to quantify the influence, 
at temperature $T$, of aging at a slightly smaller temperature $T-\Delta T$. There is 
indeed a small cumulative effect from one temperature to the other which limits the 
sharpness in temperature of the memory effect. It is this cumulative effect 
that we measure here as a function of the spin anisotropy. Secondly, we have performed 
new magnetic field change experiments in the Ising sample. This protocol enables us to 
estimate the coherence length that grows during the aging process. 
Both sets of experiments are successfully analyzed within the framework proposed in 
Ref.~\cite{Bouchaud01} (building upon ideas developed in \cite{Fisher88}, 
see also \cite{Jonsson02}), which proposes 
a real space picture for the hierarchical organization of the SG states. 
In this model, at a given temperature and time scale, aging processes are 
associated to a specific length scale. Larger length scales are frozen while smaller length 
scales are fully equilibrated. Upon reducing the temperature and as a result of the change in 
the Boltzmann weights, the small length scales which were equilibrated at higher temperature 
have to evolve towards a new equilibrium state (rejuvenation). On the other hand, the structure at large length scales can no longer evolve 
(memory). In this picture, the sharpness of the memory effect is directly related to the rapid variation of the active length scale with temperature. 
A central concern of the present paper is to establish experimentally the
time and temperature dependence of this active length scale (see Eq. \ref{exp_bouchaud} below.)

\begin{figure}
\includegraphics[width=7.5cm,height=7cm]{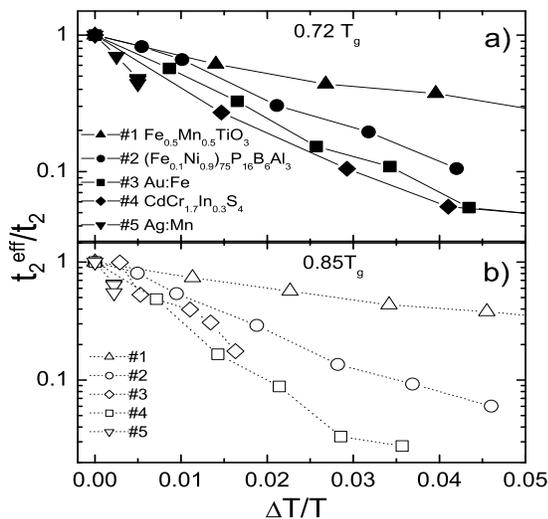}
\caption{\label{cycle} Effective waiting times deduced from the temperature cycle 
experiments at $T \simeq 0.72T_g$ (a, filled symbols) and $T \simeq 0.85T_g$ (b, open symbols). The SGs are 
labelled from \#1 to \#5 by order of decreasing anisotropy.}  
\end{figure}

\begin{figure}
\includegraphics[width=7.3cm,height=4.9cm]{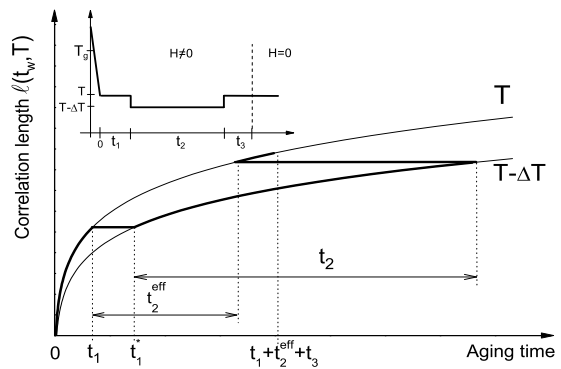}
\caption{\label{schema}
Schematic representation in terms of SG correlation length of the 
negative temperature cycling experiment depicted in the inset.}
\end{figure}

The temperature cycle experiments were performed in a monocrystalline Fe$_{0.5}$Mn$_{0.5}$TiO$_3$ Ising~\cite{Katori94} 
sample \#1 ($T_g=20.7K$) and Heisenberg SGs with decreasing random anisotropy arising from Dzyaloshinsky-Moriya interactions: 
an amorphous alloy (Fe$_{0.1}$Ni$_{0.9}$)$_{75}$P$_{16}$B$_6$Al$_3$ \#2 ($T_g=13.4K$), 
a diluted magnetic alloy Au:Fe$_{8\%}$ \#3 ($T_g=23.9K$), an insulating thiospinel 
CdCr$_{1.7}$In$_{0.3}$S$_4$ \#4 ($T_g=16.7K$) and we reproduce the data on the canonical 
Ag:Mn$_{2.7\%}$ SG \#5 ($T_g=10.4K$) from Ref.~\cite{Hammann92}. 
The relative anisotropy constants $(K/T_g)/(K/T_g)_{Ag:Mn}$, measured by torque 
experiments~\cite{Petit02}, are respectively 16.5, 8.25, 5 and 1 for samples \#2, \#3, \#4 
and \#5. An isothermal TRM procedure allows one to define a set of 
{\it reference curves} that depend on the waiting time, and to which we will
compare the relaxation curves obtained after more complicated histories. 
In our series of temperature cycle experiments, the sample 
is first cooled to a temperature $T<T_g$ under a small magnetic field (a few Oe). 
After waiting a time $t_1=500$s, the sample is further cooled to a slightly smaller 
temperature $T-\Delta T$ during $t_2=9000$s and then re-heated to $T$ for another short 
time $t_3=t_1$ (see inset in Fig.~\ref{schema}.). The additional waiting times $t_1$ and $t_3$ permit to control any possible 
thermalization delays. The magnetic field is then switched off and the TRM is recorded. 
For each `cycled' TRM curve, an effective waiting time $t_2^{eff}$ is defined such as to 
superimpose the obtained TRM with one of the purely isothermal TRM recorded at temperature $T$, 
with waiting time $t_w=t_1+t_2^{eff}(T,\Delta T)+t_3$. 
For $\Delta T=0$, 
the effective waiting time equals the actual waiting time but becomes smaller for 
$\Delta T >0$: $t_2^{eff}(T,\Delta T) < t_2 $ characterizes the effect at $T$ of aging 
at $T-\Delta T$. The results for the different SGs and for two different 
temperatures, $T/T_g \simeq 0.72$ and $T/T_g \simeq 0.85$, are plotted in Fig.~\ref{cycle} 
(for sample \#5, a similar procedure, with $t_1=t_3=0$, was applied~\cite{Hammann92}).

The slope of $t_2^{eff}/t_2$ \emph{vs} $\Delta T/T$ reflects the cumulative 
effect of aging from one temperature to the other. Hence, it quantifies the 
memory effect; the steeper the curve, the sharper the memory effect. For each of 
the temperatures $T/T_g$, \emph{the influence at $T$ of aging at $T-\Delta T$ is 
found to increase with increasing spin anisotropy}. Note that, at $T/T_g=0.72$, 
the measured $t_2^{eff}$ for sample \#3 are slightly smaller than those of the less 
anisotropic sample \#4. However, these two samples are very similar as far as their 
anisotropy is concerned since their anisotropy constants differ only by 50\% whereas 
from sample \#1 to \#5 they spread over more than one order of magnitude.

As can be seen in Fig.~\ref{fig5} below, for a given $\Delta T/T$, the effective time decreases 
with increasing temperature for all the Heisenberg-like samples (samples \#4 and \#5). By contrast, 
in the Ising sample \#1,  the influence at $T$ of a $T-\Delta T$ aging is slightly \emph{stronger} at the \emph{higher} temperature
in the Ising case. Previous \emph{ac}-measurements on the Ising sample~\cite{Dupuis01} 
demonstrated this singular behaviour even more clearly. At the higher temperature 
$T/T_g=0.84$, the dynamics was in fact found to be 
almost \emph{critical} (nearly no $\Delta T$ dependence).

To interpret these experiments in a real space picture~\cite{Bouchaud01}, 
we introduce a typical length scale $\ell(t_w,T)$ over which spin-spin correlations establish 
at temperature $T$ and waiting time $t_w$. From general arguments, this coherence length $\ell$ should be an increasing 
function of $T$ and $t_w$. For small $\Delta T$, the experimental TRM curves obtained after 
a temperature cycle superimpose onto reference isothermal curves. This implies that the SG 
has reached the same state after a $t_1+t_2^{eff}$ isothermal aging and the $t_1+t_2$ 
experimental temperature cycle. We can then sketch the temperature cycle experiment 
as in Fig.~\ref{schema}, where $t_2^{eff}$ is defined by:
\begin{equation}
\ell(t_1^*+t_2,T-\Delta T)=\ell(t_1+t_2^{eff},T)
\label{eq_cycle}
\end{equation}
where $t_1^*$ is given by $\ell(t_1,T)=\ell(t_1^*,T-\Delta T)$. Note that this picture which retains only the cumulative effect of aging with temperature (no rejuvenation effect) is only valid for small temperature cycle $\Delta T$. The slope of the experimental curve $t_2^{eff}/t_2$ \emph{vs} $(\Delta T/T)$, which quantifies the sharpness of the memory effect, reads from Eq.~(\ref{eq_cycle}) and for $\Delta T/T \ll 1$ and $t_1 \ll t_2$:  
\begin{equation}
\frac{t_2^{eff}}{t_2} \approx 
1-\left.\left ( \frac{\partial \ell / \partial \ln T}{\partial \ell / \partial \ln t_w} \right )
\right|_{t_w=t_2} \frac{\Delta T}{T}.
\label{tefft2}
\end{equation}
As announced, a sharp memory effect is related to a rapid variation of $\ell$ with the 
temperature.

\begin{figure}
\includegraphics[scale=0.62]{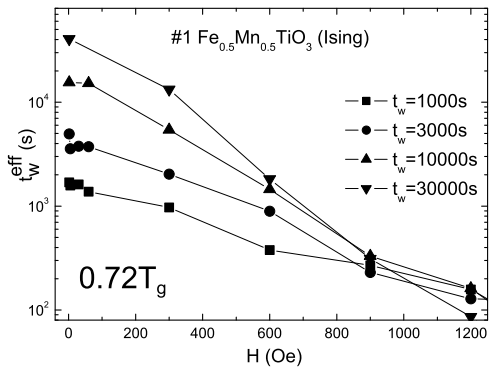}
\caption{\label{manip_champ}
Effective waiting times (log scale) derived from the field change experiments 
in the Ising sample \#1 as a function the magnetic field $H$.}
\end{figure}

In Ref.~\cite{Joh99}, magnetic field change experiments have been used to determine 
experimentally the SG coherence length in the Heisenberg samples \#4 and \#5, that was 
compared 
with numerical simulations corresponding to an Ising SG. However, the experimental determination of this length scale in the Ising case was not reported before. 
We have measured the zero field 
cooled magnetization relaxation (mirror procedure of the TRM), after an isothermal aging, 
for various amplitudes of the magnetic field $H$ applied along the $c$-axis of the Ising sample \# 1. 
The effective age $t_w^{eff}$ of the system is evaluated by a scaling procedure~\cite{Vincent95} on reference curves obtained with a small field 
($H=5$Oe). The effective times as a function of the magnetic field were measured for 
4 different waiting times and at two temperatures $0.72T_g$ and $0.92T_g$. On a semi-log scale, the data behaves rather linearly with the field (see Fig.~\ref{manip_champ}).

The effect of a field variation is to reduce the barriers that the system overcomes during aging through the Zeeman energy $E_z$. The apparent age $t_w^{eff}$ of the SG, 
smaller than the actual waiting time $t_w$, is then given, for small enough $E_z$, by:
\begin{equation}
\ln t_w^{eff}/t_w = -E_z/k_BT.
\end{equation}
The key assumption is that the Zeeman term $E_z=\mathbf{M}\cdot\mathbf{H}$ depends 
on the number $N$ of dynamically correlated spins after $t_w$. More precisely, 
for Ising spins and for not too large fields, $\mathbf{M}$ should be $\sqrt N$ times the permanent moment $m \mu_B$ of one spin, implying that $E_z$ is proportional to $H$ and $\sqrt N$ (we used a moment of $4 \mu_B$ per spins obtained from the Curie constant above $T_g$). Following Ref.~\cite{Joh99} and for the sake of simplicity, we then simply relate the number $N$ of dynamically correlated spins to the coherence length through $N=\alpha \ell^3$.

\begin{figure}
\includegraphics[width=7.5cm,height=5cm]{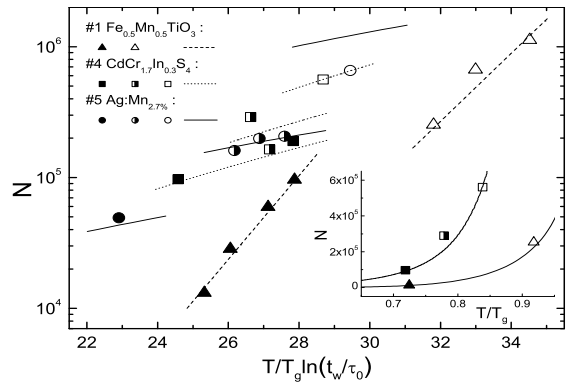}
\caption{\label{fig4} Number of correlated spins extracted from field change 
experiments in samples \#1 at $0.72T_g$ and $0.92T_g$, \#4 at $0.72T_g$, $0.78T_g$ and 
$0.84Tg$, and \#5 at $0.67T_g$, $0.77T_g$ and $0.86T_g$.  The curves are simultaneous fits
to both field variation (Eq.~\ref{exp_bouchaud}) and temperature cycling (Eq.~\ref{eq_cycle} and Eq.~\ref{exp_bouchaud}) experiments,
using a single set of parameters for each sample (table~\ref{fit_parameters}). In the main
figure, each curve segment is obtained at fixed T as a function of $t_w$ (in
this plot, the universal power law of Ref.~\cite{Joh99} is a single straight line). The
inset shows $N$ as a function of temperature after $t_w=1000$s for samples \#1 and \#4, emphasizing their different behaviors.}
\end{figure}

\begin{figure}
\includegraphics[scale=1]{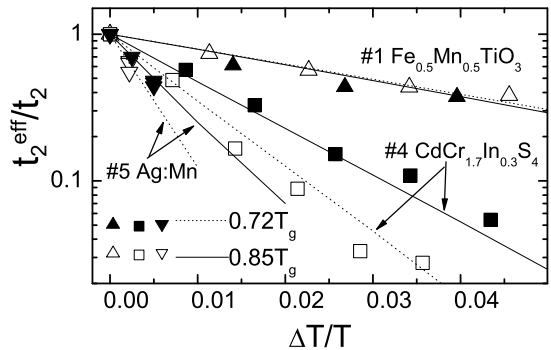}
\caption{\label{fig5} The effective waiting times deduced from the temperature cycle 
experiments are reproduced from figure~\ref{cycle}. The solid  and dashed lines are obtained from Eq.~(\ref{eq_cycle}) and Eq.~(\ref{exp_bouchaud}) with the very same parameters used 
in Fig.~\ref{fig4} (see Table~\ref{fit_parameters}).}
\end{figure}

The results of the present study along with those reproduced for the Ag:Mn 
and the thiospinel samples~\cite{Joh99} are plotted in Fig.~\ref{fig4} (in the Heisenberg case, 
$E_z \propto N$ and $N$ 
was extracted in this way in Ref.~\cite{Joh99}). For a given waiting time and 
reduced temperature, the obtained number of correlated spins in the Ising sample is 
noticeably smaller than in the Heisenberg samples studied so far, but this number grows 
faster with time, indicating a slower growth of the energy barriers with size in the 
Ising sample. The inset of Fig.~\ref{fig4} shows the evolution of $N$ with the temperature 
in samples \#1 and \#4. Clearly, the separation of the active length scales with temperature 
(``microscope effect" of the temperature~\cite{Bouchaud01}) 
is much weaker in the Ising sample. This is in complete agreement with the weak dependence 
on $\Delta T$ found in the temperature cycle experiments (small value of 
$\partial \ell / \partial \ln T$ in Eq.~(\ref{tefft2})).

\begin{table}[t]
\caption{Parameters used in Eq.~(\ref{exp_bouchaud}) to fit the
temperature cycle (Fig.~\ref{fig5}) and the magnetic
field change experiments (Fig.~\ref{fig4}). The only free parameters are $\Upsilon_0$, 
$\psi$, and $z$ since the product $z\nu$ is
known (taken from Ref.~\cite{Hammann92} and
\cite{Dupuis01}) and $\tau_0$ and $\alpha$ have been fixed (see text). The data column
gives the total number of data points that are fitted for each sample.} 
\label{fit_parameters}
%\begin{center}
\begin{ruledtabular}
\begin{tabular}{l   c   c   c   c   c   c} \\
  & $\Upsilon_0$ &  $\psi$ & $z$ & $\nu$ & z$\nu$ & data\\
\hline \\
Fe$_{0.5}$Mn$_{0.5}$TiO$_3$ & 14.5 & 0.03 & 5 & 2.1 & 10.5 & 16\\
CdCr$_{1.7}$In$_{0.3}$S$_4$ & 1.2 & 1.1 & 5.5 & 1.27 & 7 & 17\\
Ag:Mn 2.7$\%$ & 0.7 & 1.55 & 4 & 1.25 & 5 & 13\\
\end{tabular}
\end{ruledtabular}
%\end{center}
\end{table}

Now we compare quantitatively the two sets of experiments performed in this study. 
In Ref.~\cite{Joh99}, the correlation lengths were fitted by a universal power 
law  $\ell=b (t_w/\tau_0)^{a T/T_g}$ ($\tau_0$ is the microscopic attempt time and is fixed to $10^{-12}$s),
in apparent agreement with numerical simulations performed on Ising systems~\cite{Kisker96}. 
This purely activated scenario would correspond to a single straight line in Fig.~\ref{fig4}, 
and by no means accounts for our data in the Ising sample with the parameters used for the 
Heisenberg samples. Moreover, using Eq.~(\ref{tefft2}), one would find a slope $-\ln(t_2 / \tau_0)$, independent of both $a$ and $b$, for the curve 
$t_2^{eff}/t_2$ \emph{vs} $(\Delta T/T)$ shown in Fig.~\ref{cycle}.
This single value cannot reproduce the super-activated dynamics of the Heisenberg SGs 
(steep slope in Fig.~\ref{cycle}) nor the nearly critical dynamics of the Ising sample 
(weak slope in Fig.~\ref{cycle}) which requires a renormalized attempt time $\tau'_0 \gg \tau_0$. Hence, the experimental data have led us to propose 
a crossover expression~\cite{Bouchaud01,Dupuis01,Berthier02} (see also \cite{Jonsson02}):
\begin{equation}
t_w=\tau'_0 \exp \left( \Delta(T) \ell^\psi / T \right)
\label{exp_bouchaud}
\end{equation}
where $\tau'_0=\tau_0 \ell^{z}$ is a renormalized attempt time (critical dynamics) and 
$\Delta(T)=\Upsilon_0 T_g \left( 1 - T/T_g \right)^{\psi \nu}$ is the typical barrier height 
that vanishes at $T_g$ (super-activated dynamics), following Ref.~\cite{Fisher88}. 
As usual, $z$ is the dynamic critical exponent and $\nu$ is the critical exponent that governs the equilibrium correlation length. For each of the 
three SG samples \#1, \#4 and \#5, we have fitted
the whole set of experimental results to Eq.~\ref{exp_bouchaud}. These include all field
change measurements shown in Fig.~\ref{fig4} and all temperature cycling experiments
shown in Fig.~\ref{fig5} (in this case Eqs.~\ref{eq_cycle} and ~\ref{exp_bouchaud} are used). A unique set of parameters is able to account for all the
properties of each sample (see table~\ref{fit_parameters}). In order to limit the degrees of freedom in our fit, we have
imposed the $z\nu$ values, taking those derived from dynamic critical scaling
(see in refs in ~\cite{Dupuis01}). We have also fixed the geometrical factor $\alpha$ to 2
($N=\alpha \ell^3$). The parameters are not defined with a great quantitative
accuracy, since their effects on the fit are strongly correlated. However, a
consistent qualitative picture emerges. The main tendency is an
increasing value of the barrier height parameter $\Upsilon_0$ and a decreasing
barrier exponent $\psi$ for increasing values of the anisotropy. This behavior
of the exponent $\psi$ is similar to that found in the analysis of \emph{ac}~temperature cycling experiments~\cite{Dupuis01}, but it contrasts with that derived from
the scaling of $\chi''$ relaxations~\cite{Jonsson02}, as well as with that found in the
numerical simulation of Ref.~\cite{Berthier02}. Note, however, that these previous determinations were based on a 
less constrained analysis and, for the numerical work, on smaller time scales.

With a single expression for the growth of the active length, Eq.~(\ref{exp_bouchaud}), that interpolates between critical and activated 
dynamics, we have thus been able to reproduce consistently two totally different 
sets of experiments, at different temperatures, with a unique set
of parameters (for each sample). This allows us to estimate the barrier height
$\Upsilon_0$ and the barrier exponent $\psi$, which are notoriously hard to measure. As the anisotropy decreases from Ising to Heisenberg 
SGs, we find an increasingly fast separation of the active length scales with temperature, corresponding to an increased sharpness of the memory effect.
Besides, we found that the extracted 
coherence length is noticeably smaller in the Ising sample (large $\Upsilon_0$) , but grows faster with time (small $\psi$). At present, it is not clear how the strong single spin anisotropy in the Ising sample gives rise to 
both a high value of energy barriers and a very small value of the 
barrier exponent. Within a droplet description, $\psi \simeq 0$ would
imply that the droplet energy exponent $\theta$ is also zero, in agreement
with many recent numerical works on excitations in Ising SGs~\cite{Palassini03}.
Our results underline the role of anisotropy in the nature 
of a SG phase with Heisenberg spins which is presently the subject of active investigations (see \emph{e.g.} Ref.~\cite{Lee03,Petit02,Kawamura01}).

%\input{c:/users/bert/biblio/bibtex/SG.bib}
%\bibliography{c:/users/bert/biblio/bibtex/SG}% Produces the bibliography via BibTeX.

\end{document}